\documentclass[aps,prb,twocolumn,showpacs,floatfix]{revtex4}
\usepackage{graphicx}
\usepackage{latexsym}

\newcommand \be  {\begin{equation}}
\newcommand \bea {\begin{eqnarray} \nonumber}
\newcommand \ee  {\end{equation}}
\newcommand \eea {\end{eqnarray}}

\begin{document}

\title{Local excitations of a spin glass in a magnetic field}

\author{J. Lamarcq}
\author{J.-P. Bouchaud}
\author{O. C. Martin}
\altaffiliation{Laboratoire de Physique Th\'eorique et Mod\`eles Statistiques,
B\^atiment 100 --- Universit\'e Paris Sud,
91405 Orsay France}

\affiliation{Service de Physique de l'Etat Condens\'e,
Orme des Merisiers --- CEA Saclay,
91191 Gif sur Yvette Cedex France}

\begin{abstract}
We study the minimum energy clusters (\textsc{mec}) above the ground state
for the $3-d$ Edwards-Anderson Ising spin glass in a magnetic field.
For fields $B$ below $0.4$, we find that the field has almost no effect 
on the excitations that we can probe, of volume $V \leq 64$. As found 
previously for $B=0$, their energies decrease with $V$, and 
their magnetization 
remains very small (even slightly negative). For larger fields, 
both the \textsc{mec} energy and magnetization 
grow with $V$, as expected in a paramagnetic phase. However, all results 
appear to scale as $BV$ (instead of $B \sqrt{V}$ as expected 
from droplet arguments), suggesting that the spin glass phase 
is destroyed by any small field. Finally, the 
geometry of the \textsc{mec} is completely insensitive to the field, giving
further credence that they are lattice animals, in the presence or
the absence of a field.
\end{abstract}

\maketitle

\emph{Introduction ---}
Strong disorder or frustration
in a system can lead to a low temperature
phase with ``frozen order''. This is
believed to be the case for spin glasses,
the archetypes of such systems~\cite{MezardParisi87b,Young98}.
Does such a frozen
phase persist under generic perturbations? Here we
consider that question in the context of the
Edwards-Anderson~\cite{EdwardsAnderson75} Ising spin glass
when the perturbation is an external magnetic field.
The effect of the field is to
align the spins in one direction and this bias
may break up the spin glass ordering. Do arbitrarily
small fields destroy the frozen order, or can the ordering
co-exist with a field as long as it is not too large?
This is a long-standing question that has been surprisingly difficult
to answer.

From an experimental point of view, this question has been
addressed only twice for systems of Ising spins. In the first
study~\cite{ItoArugakatori94}, the (non-equilibrium) properties
were interpreted to suggest the presence of spin glass ordering at
low fields. Several years later the same sample was re-analyzed and
the lines of constant relaxation time
determined~\cite{MattssonJonsson95}. From that, a scaling analysis
was performed, leading to the conclusion that the
relaxation time is finite for non-zero fields, and thus the
system is paramagnetic. On the theoretical size, the situation remains
very controversial. In the mean field picture, where one is guided by the
Sherrington-Kirkpatrick~\cite{SherringtonKirkpatrick75} or other
mean field models, the de Almeida-Thouless
line~\cite{AlmeidaThouless78} $B_{AT}(T)$ separates the
paramagnetic and spin glass phases, and one has
$B_{AT} > 0$ if
$T < T_c$ ($T_c$ is the critical temperature in zero field).
Thus spin glass ordering continues in the presence of a
sufficiently small magnetic field in the mean field
picture. At present, there is no consensus whether this
picture correctly describes what happens in the $d=3$ Ising spin
glass, nor even when $d$ is 
large but finite. Indeed it has been argued by Fisher and
Huse~\cite{FisherHuse86} that the magnetic field is always relevant,
driving the system at large scales towards 
paramagnetism~\cite{PimentelTemesvari02}.
Then any non-zero magnetic
field will destroy the spin glass ordering and will render the system
paramagnetic beyond a length scale $\ell_B$. In the
droplet~\cite{FisherHuse86} or scaling~\cite{BrayMoore86}
pictures, $\ell_B$ diverges as $B$ vanishes like 
\be
\label{eq:magneticlength}
\ell_B \propto B^{-2/(d-2\theta)} 
\ee
where $\theta$ is the spin stiffness exponent. Because of this
divergence as $B$ decreases, the relaxation times also diverge.
The difficulty is thus to
determine whether these relaxation times diverge only as $B \to 0$
or at some positive value $B_c > 0$. Such a question is nearly
impossible to address from simulations because the time scales that can
be reached always remain quite small, while equilibrium studies
are confined to small systems for the same reason. In view of
such hurdles, much numerical 
effort~\cite{HoudayerMartin99a,KrzakalaHoudayer01}
in the last few years has
focused instead on the zero
temperature limit.

In this paper we also use zero temperature, but we study
how minimum energy clusters~\cite{LamarcqBouchaud02} above
the ground state are affected by a magnetic field.
Essentially, we search for a putative critical field $B_c$; 
if there is an AT-line, $B_c$ is then simply
$B_{AT}(T=0)$. Our investigation reduces
to finding whether there is a phase transition in $B$ at
$T=0$, and for that 
we need an order parameter.
Since the field breaks the up-down symmetry,
the Edwards-Anderson order parameter that gives the spatial
average of the square magnetization at equilibrium is
not of use. It is thus necessary to characterize the
spin glass ordering otherwise, for instance
by an infinite (spin glass) susceptibility or by the presence of
irreversibility on all time scales. However these quantities
are not accessible at $T=0$, and so other measures of
ordering are necessary. One characteristic of a spin glass
order is sensitivity to external perturbations; one can thus
see whether the ground state evolves chaotically with $B$
up to a critical value $B_c$. The study in~\cite{HoudayerMartin99a}
showed that there was no sign of chaos when the lattice size
was increased for values of $B\ge 1$, suggesting that
$0 \le B_c < 1.0$. This should be compared to the mean field value
for connectivity $6$ random graphs for which $B_{AT}(T=0) \simeq 2.1$.
Another study~\cite{KrzakalaHoudayer01}
considered how the ground state responded to a domain wall twist;
this gave some evidence that the system had frozen order for
$B \le B_c\simeq 0.6$, although $B_c=0$ could not be excluded.
Our approach here probes the spin glass ordering through its
{\em local} excitations; the energy to flip a cluster of spins
behaves differently in a spin glass phase and in a
paramagnetic phase, and we use this to test for spin glass ordering.
First we shall describe what these minimum energy clusters are;
then we shall consider how their energies and magnetizations
depend on the external magnetic field.

\emph{Minimum energy clusters ---}
We start with the $3-d$ Edwards-Anderson (EA) model with periodic
boundary conditions. The Hamiltonian is defined on a cubic lattice
of $N=L^3$ spins,
\be \label{eq:EA} H =-\sum_{<ij>} J_{ij} S_i S_j
+ B \sum_i S_i \ . 
\ee
The spins are Ising, i.e., $S_i = \pm 1$,
and the nearest-neighbor interactions $\{J_{ij}\}$ are quenched
random variables distributed according to a Gaussian law with zero
mean and unit variance.

In this model, when $B$ is set to $0$, there is numerical
evidence~\cite{PalassiniCaracciolo99} of an ordered phase at low temperature
where each spin $i$ is frozen in a random 
direction $\langle S_i \rangle \neq 0$.
We are interested in local excitations in the putative frozen phase,
so following previous work~\cite{LamarcqBouchaud02},
we define a Minimum Energy Cluster or \textsc{mec} as 
the {\em connected} cluster
of spins of lowest energy that contains a specified number of spins and
a given, arbitrarily chosen site. The study of such \textsc{mec} in 
{\em zero field}
showed~\cite{LamarcqBouchaud02} that they are, perhaps surprisingly, fractals
whose typical energy {\em decreases} with their size, in contrast with 
the properties expected for the Fisher-Huse droplets. This result is striking and we take it to be a signature of 
the spin glass phase, allowing us to probe now the case where $B > 0$.

\emph{Numerical method ---}
Our measurements are performed on lattices with $L=6$ and
$L=10$ with
different values of the field taken in the range $0 \le B \le 1.5$; in
all cases, we generated $100$ disorder samples which was
enough to obtain reasonably small statistical errors on our observables.

For each disorder sample, we first compute the ground state of the
system using a genetic renormalization algorithm~\cite{HoudayerMartin99b}.
Then, we choose an arbitrary ``reference'' spin and flip it
along with a cluster containing $V-1$ other spins connected
to it.  The goal is to find the \textsc{mec} of size
$V$, with the constraint that the reference spin is held flipped and
the cluster is always connected. To find that \textsc{mec},
we use non-local Kawasaki dynamics as in \cite{LamarcqBouchaud02}
and exchange Monte Carlo~\cite{HukushimaNemoto96} with
a set of 30 temperatures (between $T_1=0.5$ and $T_{30}=3.0$).
This search for the \textsc{mec} is repeated five
independent times to check whether the same cluster is found.
(Since our algorithm is
heuristic, we obtain an upper bound rather than the
true energy of the \textsc{mec}.) In practice, we find
the \textsc{mec} quite
reliably for $V \le 32$ as shown by our tests.
Note that our optimization
procedure has been improved since the analysis made
in~\cite{LamarcqBouchaud02} in zero field; for instance, we have
been able to find, for $V=64$, lower excitations than before.
With this improvement, we confirm our earlier conclusions but also 
we have been able to go to larger \textsc{mec} sizes.
In all that follows, we take
$V_n=2^n,\ 1 \le n \le 6$; even though the true 
\textsc{mec} are not always found at $V=64$, we present
the results also for that size as we believe
the corresponding bias is small.

\begin{figure}[htb]
\begin{center}
\includegraphics[width=7.5cm,height=5cm]{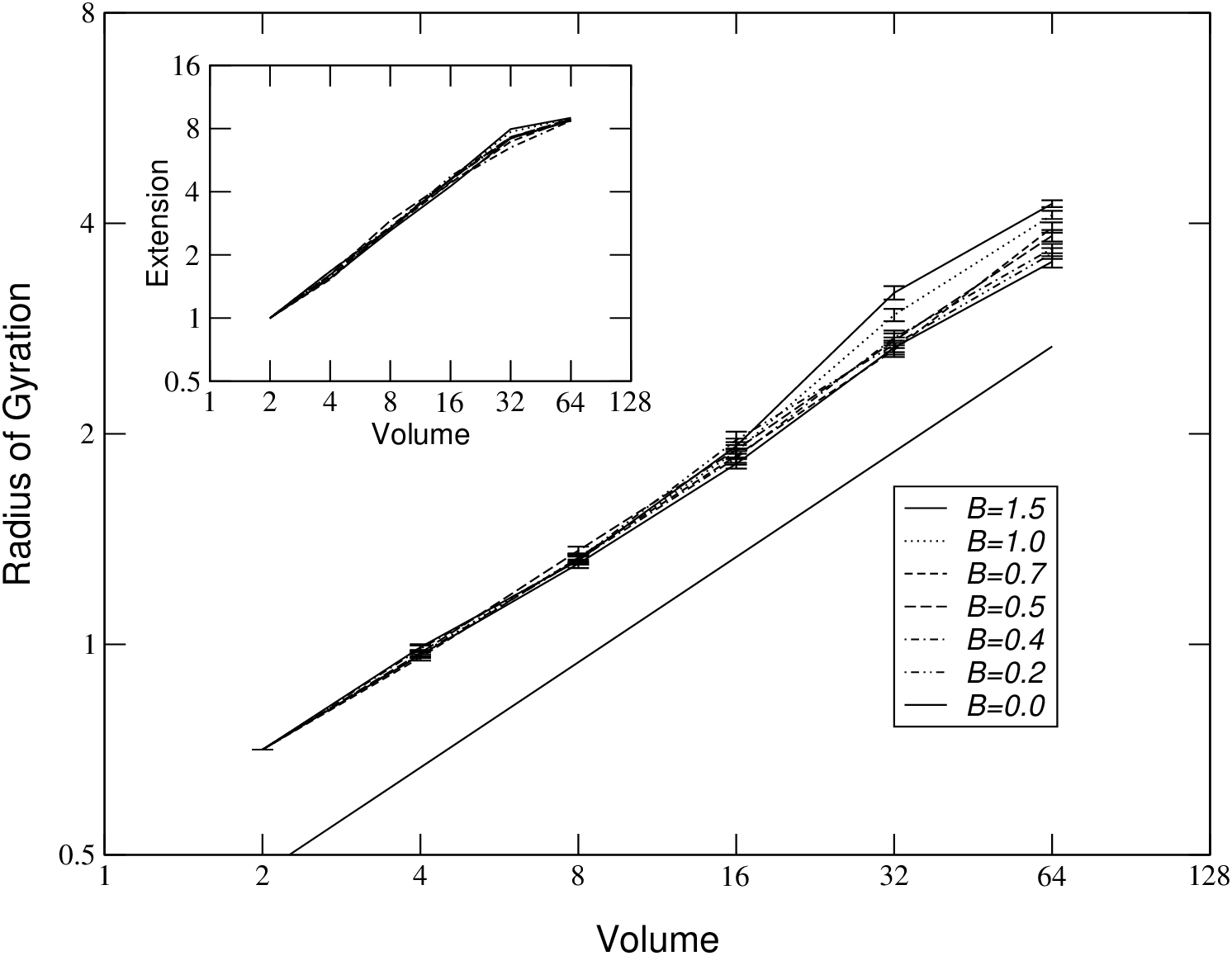}
\vspace{-1mm}
\caption{\label{fig:geometry}Radius of gyration versus the
volume of \textsc{mec} at different values of the field $B$ 
when $L=10$. Insert: Mean end-to-end extension versus the volume.
Note the log-log scale in both cases. The plain line has a slope $1/2$,
corresponding to $d_f=2$.}
\vspace{-5mm}
\end{center}
\end{figure}

\emph{Geometry of \textsc{mec} ---}
Let us look at two geometric properties of \textsc{mec}.
Consider first the
extension of \textsc{mec} as a function of their volume $V$;
our data for the mean end-to-end extensions are displayed in
the insert of 
Fig.~\ref{fig:geometry}. Given these values, we 
conclude that \textsc{mec} span the whole system
when $V \approx 15$ if $L=6$ and when $V \approx 30$ if $L=10$.

Second, consider the fractal dimension
of the \textsc{mec}. When $B=0$, we argued~\cite{LamarcqBouchaud02}
that configurational entropy is sufficient to drive \textsc{mec} towards
the lattice animal phase in which they are stringy objects with a
fractal dimension $d_f = 2$. This excitation branch is thus distinct 
from the Fisher-Huse droplets which are compact objects with a fractal 
{\em boundary}. What happens in the
presence of a field? It is useful to think of the
large $B$ limit which is strongly paramagnetic.
The surface energy of a cluster is a random variable $J_{ij}$
that is symmetric about zero,
while its bulk energy is $\approx BV$.
Clearly, energy minimization in a random environment 
will drive the clusters to extend their surface,
leading to the lattice animal phase. We can see in
Fig.~\ref{fig:geometry} that this geometrical property,
just like the extension, is nearly independent
of the strength of the magnetic field. This shows that geometrical 
properties of \textsc{mec} are {\em not} relevant to
distinguish between a paramagnet and a spin glass, but it also gives
further support to the claim that \textsc{mec} are lattice
animals even when $B=0$.

\emph{Energies of \textsc{mec} ---}
Consider first the typical excitation energy $\overline{E}$ of
a \textsc{mec}, where $(\overline{\cdot})$ denotes 
the average over disorder. 
In the paramagnetic phase, i.e., for large $B$,
$\overline{E}$ is expected to be self-averaging as $V\to \infty$. Since there
is a non zero magnetization per spin, that energy should grow as
$\overline{E(V)} \propto BV$, leading therefore to $\theta = 3$. Note however
that because of the self-averaging property, the probability to find
$E = O(1)$ should go to zero much faster than
any power of $V$ for large $V$; there are thus {\em no}
low-energy excitations at large scales.
Following an Imry-Ma argument, Fisher and Huse~\cite{FisherHuse86}
argued that this should actually occur for any $B\ne0$ for sufficiently 
large $V$'s. The paramagnetic behavior should set in beyond a 
cross-over volume $V_B$,
obtained by comparing the Zeeman energy $B\sqrt{V}$ of a $B=0$ 
droplet with its 
excitation energy $\Upsilon V^{\theta/d_f}$. The result is
the cross-over volume $V_B \propto B^{-2d_f/(d_f-2\theta)}$,
with the corresponding length being given in Eq.~\ref{eq:magneticlength}. 
Note that
$d_f$ is the (possibly fractal) dimension of the excitation.
In the usual picture of compact droplets~\cite{FisherHuse87},
one has $d_f=3$ and $\theta \approx 0.2$. In the numerical
investigation of \textsc{mec}~\cite{LamarcqBouchaud02},
it was found that the excitation clusters were instead fractal,
$d_f \approx 2.0$, and that the effective $\theta$ was small and
negative. Since $\theta$ is small 
compared to $d_f$ in all cases, we expect roughly
$V_B \propto (\Upsilon/B)^{2}$. Using the value $\Upsilon \approx 6$ from our
numerical results, we obtain a rather large value $V_B \approx 36/B^2$. 

\begin{figure}[htb]
\begin{center}
\includegraphics[width=7.5cm,height=5cm]{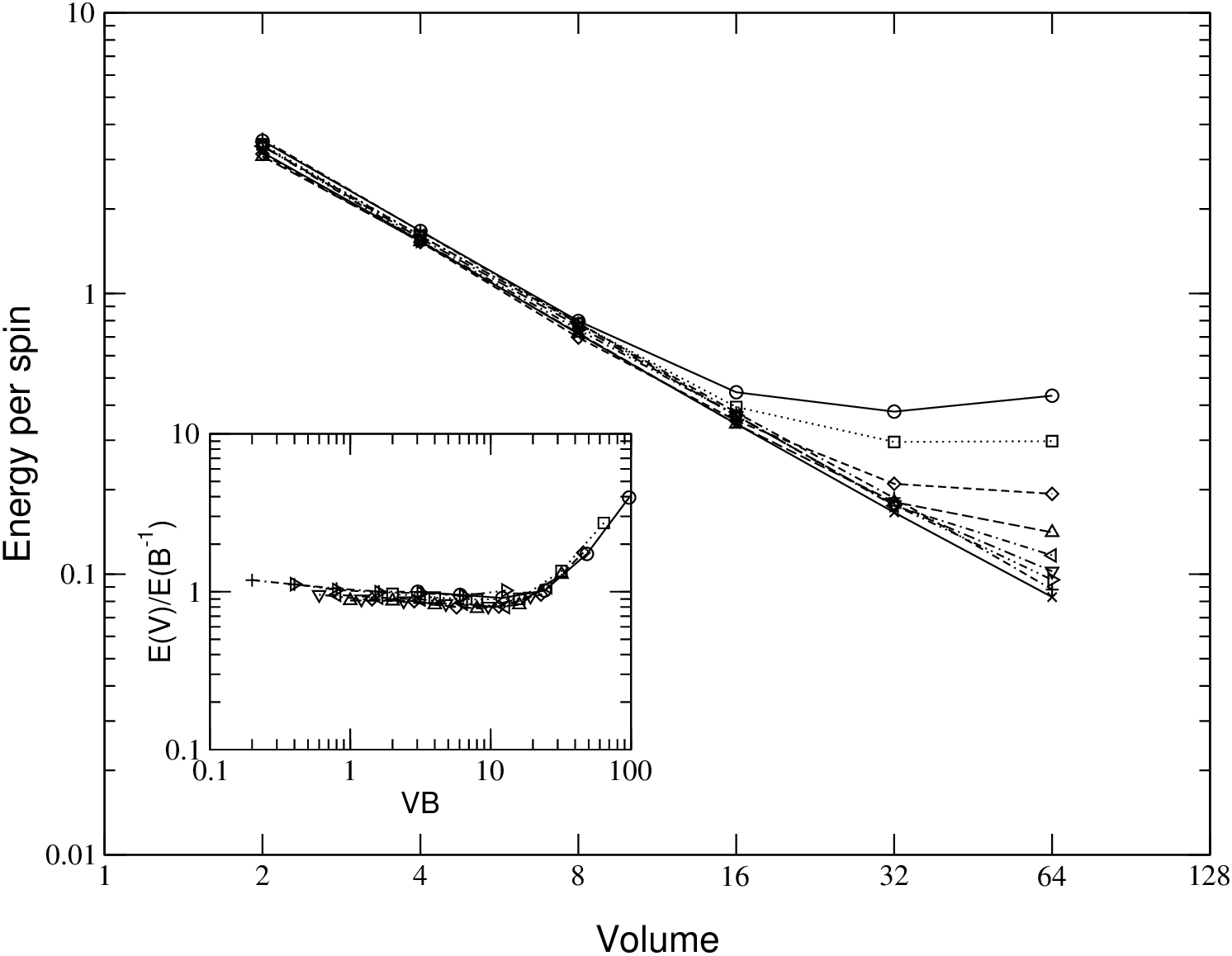}
\vspace{-1mm}
\caption{\label{fig:energies}Mean energy per spin versus volume at
different values of $B$ when $L=10$. Insert: Scaling curve.
$B=1.5(\circ)$, $1.0(\Box)$, $0.7(\Diamond)$, 
$0.5(\bigtriangleup)$, $0.4(\lhd)$,
$0.3(\bigtriangledown)$, $0.2(\rhd)$, $0.1(+)$, $0(\times)$}
\vspace{-5mm}
\end{center}
\end{figure}

Let us focus on the volume dependence of the mean \textsc{mec} energy at
different fields $B$. For all our $B$ values, this mean
initially {\em decreases} with volume; furthermore, this decay
is compatible with a power law, $\overline{E(V)} \approx V^{\theta_f/d_f}$
with $\theta_f \approx -0.13$ as found in zero 
field~\cite{LamarcqBouchaud02}. From the point of view of the mean 
field picture, the behavior observed in our data
is qualitatively as expected: below $B_c \approx 0.5$,
the spin glass ordering leads to a decreasing
$\overline{ E(V)}$ at all $V$, while for $B>B_c$ we recover the paramagnetic
behavior of an increasing energy when $V$ is sufficiently large.
Note however that the estimate $B_c \approx 0.5$
is far below the mean field value~\cite{HoudayerMartin99a} associated with
random graphs of connectivity $6$, $B_{MF} \approx 2.1$. The important issue 
is therefore whether the curves eventually bend upwards
at large $V$ for all values of $B>0$. In order to be more quantitative, 
we have attempted to scale 
the data, looking for a collapse when plotting
$\overline{ E(V)} / \overline{ E(V_B)}$ as a function of $V/V_B$. 
Surprisingly, the data does not collapse well at all when $V_B \propto B^{-2}$ 
but merges much better if we take $V_B \propto B^{-1}$ as shown
by the ``scaling curve'' 
in the insert of Fig.~\ref{fig:energies}. This result suggests 
that even the low $B$
curves might eventually bend upwards for large enough $V$;
that could be interpreted 
as signaling the instability of the spin-glass phase in 
a field. We will return to this
point later.

\begin{figure}[htb]
\begin{center}
\includegraphics[width=7.5cm,height=5cm]{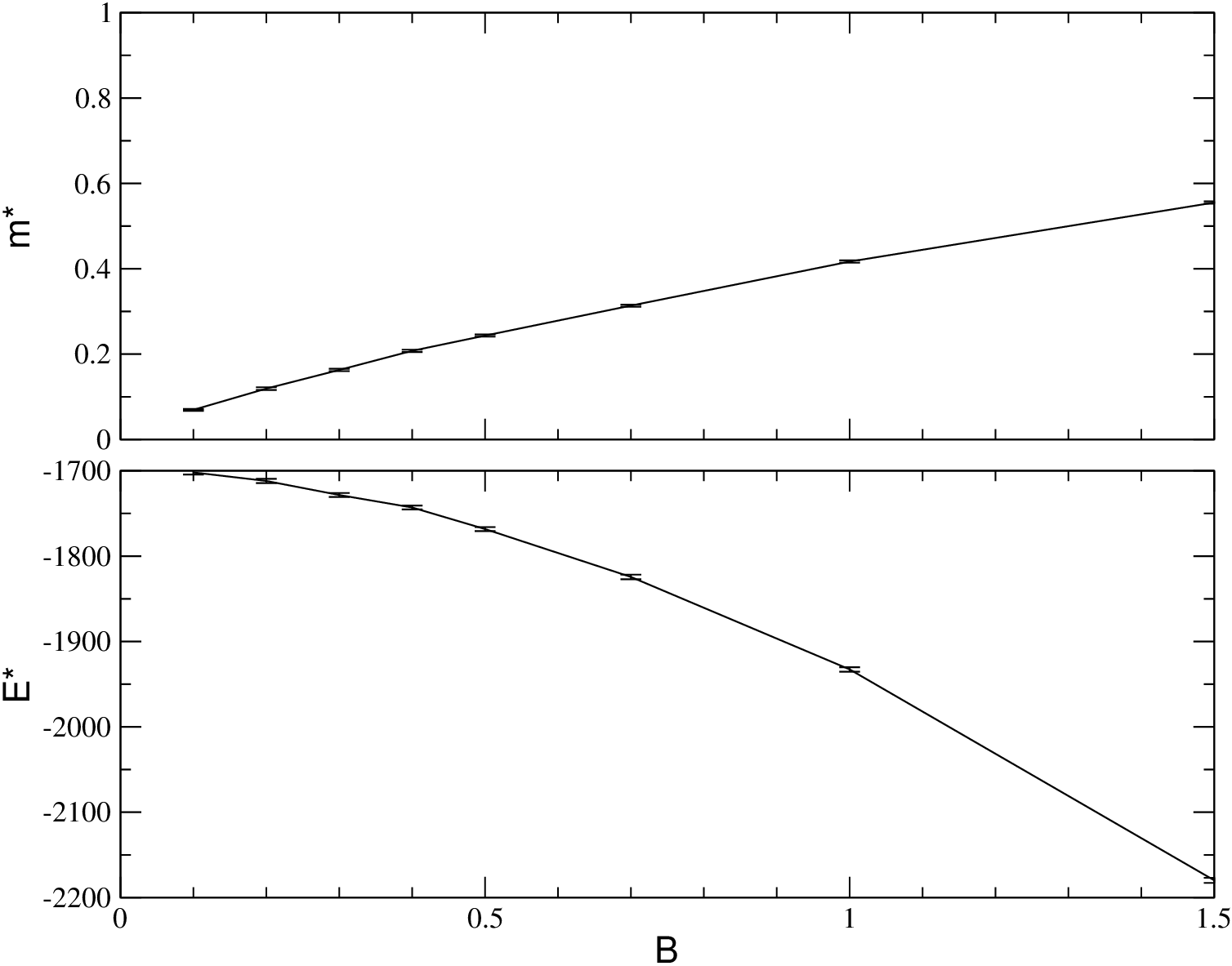}
\vspace{-1mm}
\caption{\label{fig:ground}Mean magnetization per site (upper panel) and 
mean energy (lower panel) of the ground state of a $N=10^3$ spins 
Edwards-Anderson system, versus the magnetic field.}
\vspace{-5mm}
\end{center}
\end{figure}

Note that the $BV$ scaling obtained above is rather unexpected 
since it means that
the influence of a small magnetic field on low energy excitations 
is actually {\em stronger} than anticipated by the Fisher-Huse 
argument. Although 
we do not have a clear understanding of this result, one could slightly 
alter the argument as follows. 
For $B=0$ the ground state has a zero magnetization per 
spin, which means, as recalled 
above, that the magnetization of a region of volume $\ell^3$ is 
typically of order 
$\ell^{3/2}$. For $B \neq 0$, a non zero magnetization per spin $\chi B$ 
(see Fig.~\ref{fig:ground}) appears; this
means that the spatial correlations {\em in the ground state} change for 
distances larger
than $\ell_B$ such that $\ell_B^3 \chi B \approx \ell_B^{3/2}$, leading to
$\ell_B \approx (\chi B)^{-2/3}$. Since the \textsc{mec} are fractal, 
the excitations built at
$B \neq 0$ are expected to be affected as soon as $V \sim \ell_B^{d_f}$, or, 
using $d_f=2$, at $V \sim (\chi B)^{-4/3}$, a result that
is closer to the scaling reported in 
Fig.~\ref{fig:energies}.

\begin{figure}[htb]
\begin{center}
\includegraphics[width=7.5cm,height=5cm]{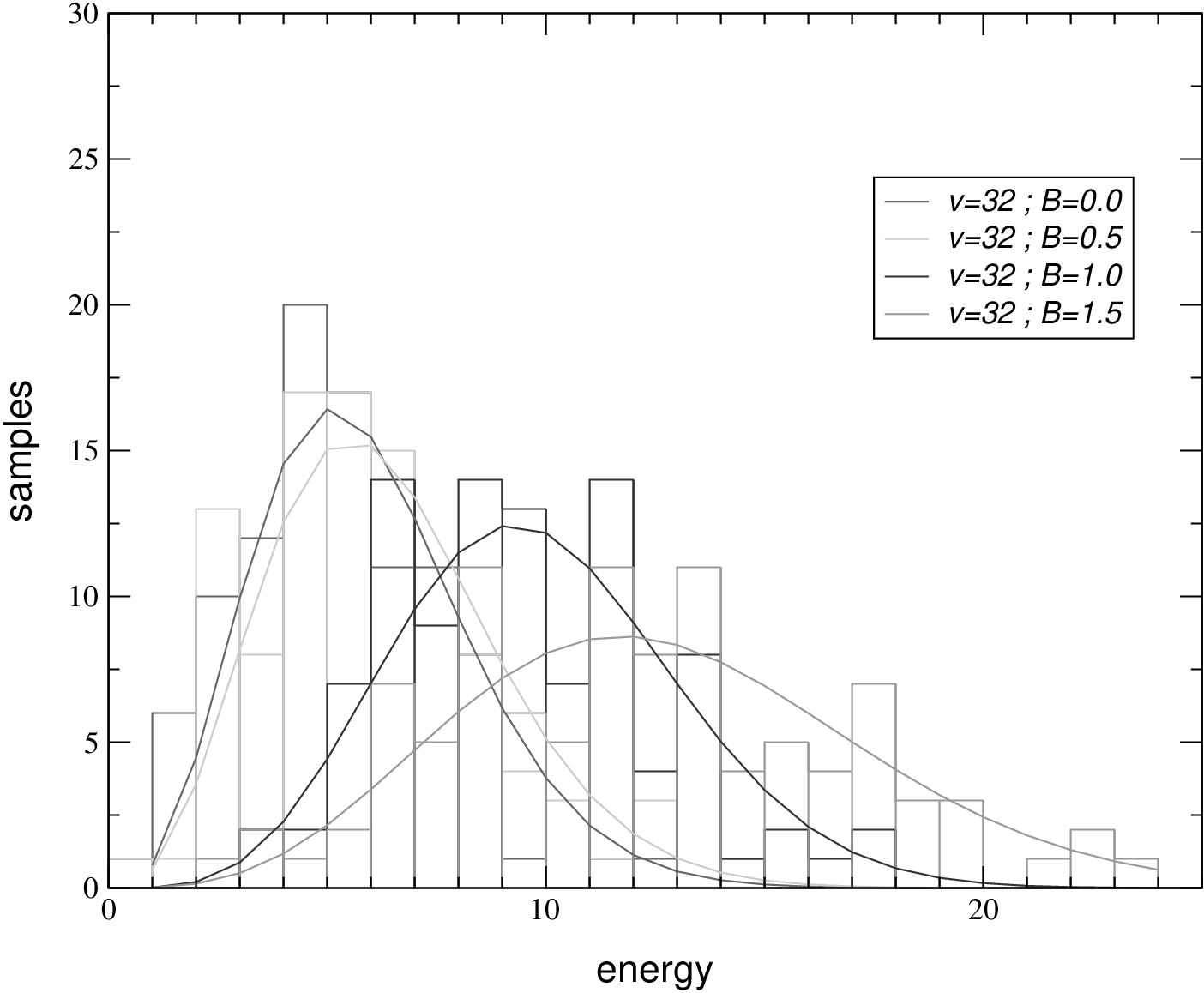}
\vspace{-1mm}
\caption{\label{fig:histos}Histogram of energies for $V=32$ \textsc{mec}, at
different values of $B$. The lines are there to guide the eye.}
\vspace{-5mm}
\end{center}
\end{figure}

We have also studied the {\em distribution} of energies
of \textsc{mec}, for a given volume; the case
$V=32$ is given in Fig.~\ref{fig:histos}. As expected, for 
large fields ($B=1.0$ and $1.5$), the distributions become
symmetric around their maximum, and their weights for small
excitation energies rapidly tend to zero. 
However for low fields, $B \le 0.5$, we find that 
these distributions hardly
change at all with $B$.

\begin{figure}[htb]
\begin{center}
\vspace{0mm}
\includegraphics[width=7.5cm,height=5cm]{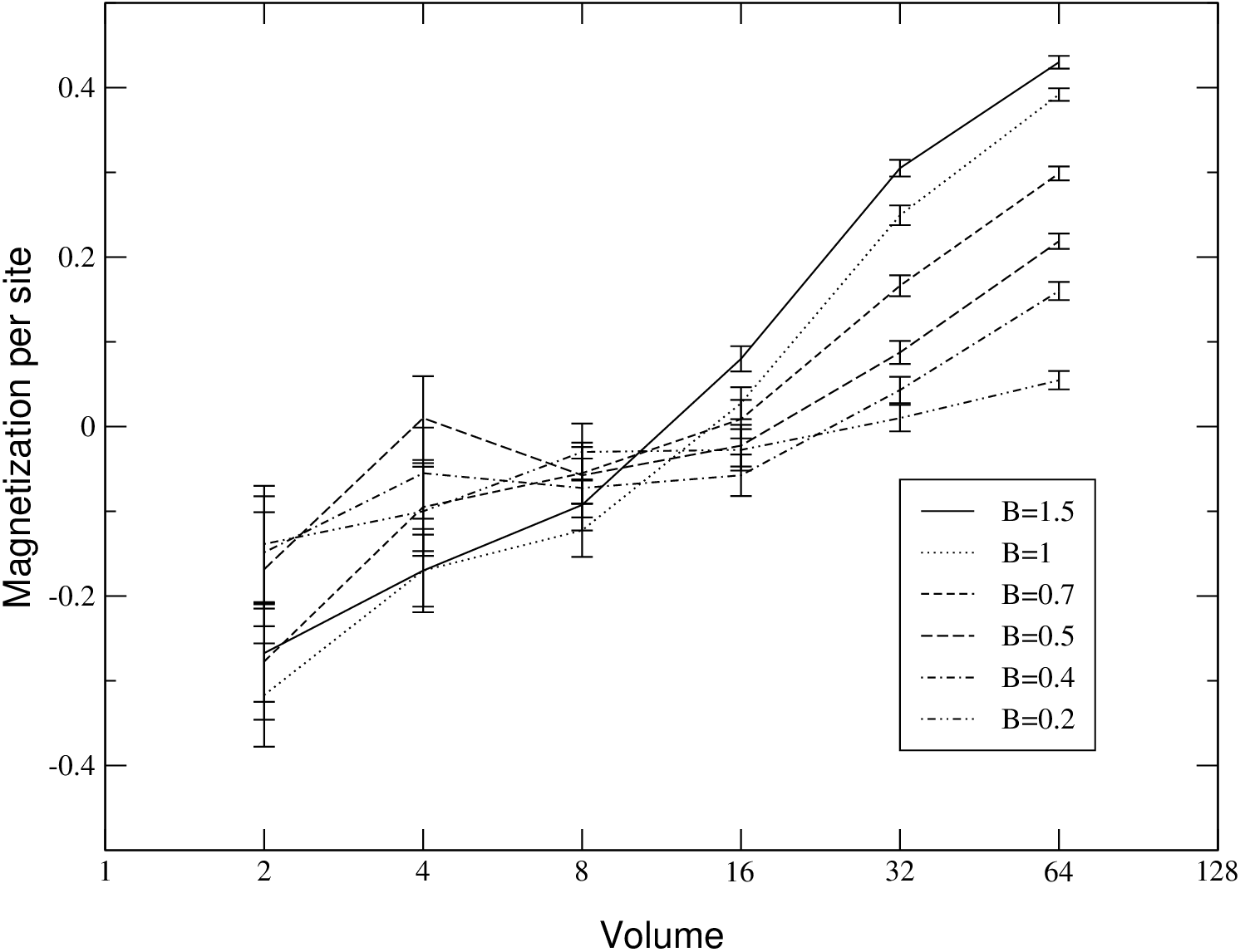}
\caption{\label{fig:magnetization}Mean magnetization per site 
versus $V$ when $L=10$, for different magnetic fields $B$.}
\vspace{-8mm}
\end{center}
\end{figure}

\emph{Magnetization of \textsc{mec} ---}
Another observable of interest is the
mean magnetization per site $m(V,B)$ of \textsc{mec} of volume $V$
in the presence of the magnetic field $B$. (The
\textsc{mec}'s magnetization is defined {\em before} it is flipped.) 
In the droplet picture, the system is driven to the large $B$ limit 
when $V$ grows
as soon as $B \ne 0$. Thus, beyond the cross-over scale $V_B$ discussed above,
$m(V,B)$ should converge to a function $m(B)$ that is non-zero
for all $B \ne 0$. The situation is different in the mean field
picture. Mean field predicts that the
equilibrium magnetization is independent of the temperature
$T$ at low enough $T$. Now since the 
low $T$ properties follow from those of the lowest-lying excitations, 
we see that these excitations cannot have a non-zero
mean magnetization. The mean field picture then consists in
having the large $V$ limit of $m(V,B)$ vanish for $B \le B_c$ and
grow for $B>B_c$. In both pictures of the spin glass phase, $m(B)$ acts as an 
{\em order parameter} for the paramagnetic phase.
Fig.~\ref{fig:magnetization} shows $m(V,B)$ as a function of $V$.
We see that the magnetizations for $B \le 0.5$ are indeed very
small and in fact are negative rather than positive. Also,
we see a clear upturn to positive $m$ for $V=32$, $B \gtrsim 0.4$; this upturn 
seems to be present even for smaller fields, suggesting that $B_c < 0.4$. This bound is more stringent than what we found from the {\em energies}
of the \textsc{mec}. Finally, just like the energy, the magnetization 
appears to approximately rescale as a function of $BV$. 

\emph{Conclusions ---}
Our results show that when $B < 0.4$, the system behaves qualitatively
just as in the case $B=0$ (at least for the sizes studied here), 
whereas significant changes
arise when $B \ge 0.4$. If there is a critical field, which appears to us
rather unplausible, it 
must obey $B_c < 0.4$, 
a bound that is stronger than that most recently 
obtained~\cite{KrzakalaHoudayer01}.
This value is small compared to the value arising
in the mean field approximation for which $B_c \approx 2.1$. Although 
we cannot 
completely rule out the mean field picture, the possibility of rescaling the 
different curves as a function of $BV$ points towards the destruction of the
spin-glass phase for $B \neq 0$. On the other hand,
our data are also unexpected from the point of view of the
droplet model. First, as emphasized before~\cite{LamarcqBouchaud02}, {\sc mec} 
appear as fractal, rather than compact, objects. Second, the Fisher-Huse 
prediction of a crossover volume $V_B \propto B^{-2}$ appears not to be
correct. We find that the magnetic field has a much stronger effect, since we 
obtain $V_B \propto B^{-1}$, which might be a consequence of the 
fractal nature 
of the {\sc mec}. Finally, the geometric properties of 
the {\sc mec} appear to
be $B$ independent, and most probably they are
lattice animals for all $B$. Many of the points raised by the present 
study would be clarified if the corresponding results in $d=4$ were available.
Note that recent work~\cite{KrzakalaParisi03} on
Bethe lattices reveals
a non zero value of the critical field $B_c$ as expected.

\emph{Acknowledgments ---}
J. L. acknowledges a fellowship from the MENRT and the CEA for
computer time on a Compaq SC232. O. M. thanks the SPEC
for its hospitality during this work. We thank F. Krzakala for 
communicating his results prior to publication.

\bibliographystyle{apsrev}
\bibliography{../../../Bib/references}

\addcontentsline{toc}{chapter}{\protect\bibname}
\begin{thebibliography}{18}
\expandafter\ifx\csname natexlab\endcsname\relax\def\natexlab#1{#1}\fi
\expandafter\ifx\csname bibnamefont\endcsname\relax
  \def\bibnamefont#1{#1}\fi
\expandafter\ifx\csname bibfnamefont\endcsname\relax
  \def\bibfnamefont#1{#1}\fi
\expandafter\ifx\csname citenamefont\endcsname\relax
  \def\citenamefont#1{#1}\fi
\expandafter\ifx\csname url\endcsname\relax
  \def\url#1{\texttt{#1}}\fi
\expandafter\ifx\csname urlprefix\endcsname\relax\def\urlprefix{URL }\fi
\providecommand{\bibinfo}[2]{#2}
\providecommand{\eprint}[2][]{\url{#2}}

\bibitem[{\citenamefont{M{\'e}zard et~al.}(1987)\citenamefont{M{\'e}zard,
  Parisi, and Virasoro}}]{MezardParisi87b}
\bibinfo{author}{\bibfnamefont{M.}~\bibnamefont{M{\'e}zard}},
  \bibinfo{author}{\bibfnamefont{G.}~\bibnamefont{Parisi}}, \bibnamefont{and}
  \bibinfo{author}{\bibfnamefont{M.~A.} \bibnamefont{Virasoro}},
  \emph{\bibinfo{title}{Spin-Glass Theory and Beyond}},
  vol.~\bibinfo{volume}{9} of \emph{\bibinfo{series}{Lecture Notes in Physics}}
  (\bibinfo{publisher}{World Scientific}, \bibinfo{address}{Singapore},
  \bibinfo{year}{1987}).

\bibitem[{\citenamefont{Young}(1998)}]{Young98}
\bibinfo{editor}{\bibfnamefont{A.~P.} \bibnamefont{Young}}, ed.,
  \emph{\bibinfo{title}{Spin Glasses and Random Fields}}
  (\bibinfo{publisher}{World Scientific}, \bibinfo{address}{Singapore},
  \bibinfo{year}{1998}).

\bibitem[{\citenamefont{Edwards and Anderson}(1975)}]{EdwardsAnderson75}
\bibinfo{author}{\bibfnamefont{S.~F.} \bibnamefont{Edwards}} \bibnamefont{and}
  \bibinfo{author}{\bibfnamefont{P.~W.} \bibnamefont{Anderson}},
  \bibinfo{journal}{J. Phys. F} \textbf{\bibinfo{volume}{5}},
  \bibinfo{pages}{965} (\bibinfo{year}{1975}).

\bibitem[{\citenamefont{Ito and Katori}(1994)}]{ItoArugakatori94}
\bibinfo{author}{\bibfnamefont{A.}~\bibnamefont{Ito}} \bibnamefont{and}
  \bibinfo{author}{\bibfnamefont{H.~A.} \bibnamefont{Katori}},
  \bibinfo{journal}{J. Soc. Japan} \textbf{\bibinfo{volume}{63}},
  \bibinfo{pages}{3122} (\bibinfo{year}{1994}).

\bibitem[{\citenamefont{Mattsson et~al.}(1995)\citenamefont{Mattsson, Jonsson,
  Nordblad, Katori, and Ito}}]{MattssonJonsson95}
\bibinfo{author}{\bibfnamefont{J.}~\bibnamefont{Mattsson}},
  \bibinfo{author}{\bibfnamefont{T.}~\bibnamefont{Jonsson}},
  \bibinfo{author}{\bibfnamefont{P.}~\bibnamefont{Nordblad}},
  \bibinfo{author}{\bibfnamefont{H.~A.} \bibnamefont{Katori}},
  \bibnamefont{and} \bibinfo{author}{\bibfnamefont{A.}~\bibnamefont{Ito}},
  \bibinfo{journal}{Phys. Rev. Lett.} \textbf{\bibinfo{volume}{74}},
  \bibinfo{pages}{4305} (\bibinfo{year}{1995}).

\bibitem[{\citenamefont{Sherrington and
  Kirkpatrick}(1975)}]{SherringtonKirkpatrick75}
\bibinfo{author}{\bibfnamefont{D.}~\bibnamefont{Sherrington}} \bibnamefont{and}
  \bibinfo{author}{\bibfnamefont{S.}~\bibnamefont{Kirkpatrick}},
  \bibinfo{journal}{Phys. Rev. Lett.} \textbf{\bibinfo{volume}{35}},
  \bibinfo{pages}{1792} (\bibinfo{year}{1975}).

\bibitem[{\citenamefont{de~Almeida and Thouless}(1978)}]{AlmeidaThouless78}
\bibinfo{author}{\bibfnamefont{J.~R.~L.} \bibnamefont{de~Almeida}}
  \bibnamefont{and} \bibinfo{author}{\bibfnamefont{D.~J.}
  \bibnamefont{Thouless}}, \bibinfo{journal}{J. Phys. A}
  \textbf{\bibinfo{volume}{11}}, \bibinfo{pages}{983} (\bibinfo{year}{1978}).

\bibitem[{\citenamefont{Fisher and Huse}(1986)}]{FisherHuse86}
\bibinfo{author}{\bibfnamefont{D.~S.} \bibnamefont{Fisher}} \bibnamefont{and}
  \bibinfo{author}{\bibfnamefont{D.~A.} \bibnamefont{Huse}},
  \bibinfo{journal}{Phys. Rev. Lett.} \textbf{\bibinfo{volume}{56}},
  \bibinfo{pages}{1601} (\bibinfo{year}{1986}).

\bibitem[{\citenamefont{Pimentel et~al.}(2002)\citenamefont{Pimentel,
  Temesvari, and Dominicis}}]{PimentelTemesvari02}
\bibinfo{author}{\bibfnamefont{I.~R.} \bibnamefont{Pimentel}},
  \bibinfo{author}{\bibfnamefont{T.}~\bibnamefont{Temesvari}},
  \bibnamefont{and} \bibinfo{author}{\bibfnamefont{C.~D.}
  \bibnamefont{Dominicis}}, \bibinfo{journal}{Phys. Rev. B}
  \textbf{\bibinfo{volume}{65}}, \bibinfo{pages}{224420}
  (\bibinfo{year}{2002}).

\bibitem[{\citenamefont{Bray and Moore}(1986)}]{BrayMoore86}
\bibinfo{author}{\bibfnamefont{A.~J.} \bibnamefont{Bray}} \bibnamefont{and}
  \bibinfo{author}{\bibfnamefont{M.~A.} \bibnamefont{Moore}}, in
  \emph{\bibinfo{booktitle}{Heidelberg Colloquium on Glassy Dynamics}}, edited
  by \bibinfo{editor}{\bibfnamefont{J.~L.} \bibnamefont{van Hemmen}}
  \bibnamefont{and}
  \bibinfo{editor}{\bibfnamefont{I.}~\bibnamefont{Morgenstern}}
  (\bibinfo{publisher}{{S}pringer}, \bibinfo{address}{{B}erlin},
  \bibinfo{year}{1986}), vol. \bibinfo{volume}{275} of
  \emph{\bibinfo{series}{Lecture Notes in Physics}}, pp.
  \bibinfo{pages}{121--153}.

\bibitem[{\citenamefont{Houdayer and
  Martin}(1999{\natexlab{a}})}]{HoudayerMartin99a}
\bibinfo{author}{\bibfnamefont{J.}~\bibnamefont{Houdayer}} \bibnamefont{and}
  \bibinfo{author}{\bibfnamefont{O.~C.} \bibnamefont{Martin}},
  \bibinfo{journal}{Phys. Rev. Lett.} \textbf{\bibinfo{volume}{82}},
  \bibinfo{pages}{4934} (\bibinfo{year}{1999}{\natexlab{a}}),
  \bibinfo{note}{cond-mat/9811419}.

\bibitem[{\citenamefont{Krzakala et~al.}(2001)\citenamefont{Krzakala, Houdayer,
  Marinari, Martin, and Parisi}}]{KrzakalaHoudayer01}
\bibinfo{author}{\bibfnamefont{F.}~\bibnamefont{Krzakala}},
  \bibinfo{author}{\bibfnamefont{J.}~\bibnamefont{Houdayer}},
  \bibinfo{author}{\bibfnamefont{E.}~\bibnamefont{Marinari}},
  \bibinfo{author}{\bibfnamefont{O.~C.} \bibnamefont{Martin}},
  \bibnamefont{and} \bibinfo{author}{\bibfnamefont{G.}~\bibnamefont{Parisi}},
  \bibinfo{journal}{Phys. Rev. Lett.} \textbf{\bibinfo{volume}{87}},
  \bibinfo{pages}{197204} (\bibinfo{year}{2001}),
  \bibinfo{note}{cond-mat/0107366}.

\bibitem[{\citenamefont{Lamarcq et~al.}(2002)\citenamefont{Lamarcq, Bouchaud,
  Martin, and M\'ezard}}]{LamarcqBouchaud02}
\bibinfo{author}{\bibfnamefont{J.}~\bibnamefont{Lamarcq}},
  \bibinfo{author}{\bibfnamefont{J.-P.} \bibnamefont{Bouchaud}},
  \bibinfo{author}{\bibfnamefont{O.~C.} \bibnamefont{Martin}},
  \bibnamefont{and} \bibinfo{author}{\bibfnamefont{M.}~\bibnamefont{M\'ezard}},
  \bibinfo{journal}{Europhys. Lett.} \textbf{\bibinfo{volume}{58}},
  \bibinfo{pages}{321} (\bibinfo{year}{2002}).

\bibitem[{\citenamefont{Palassini and
  Caracciolo}(1999)}]{PalassiniCaracciolo99}
\bibinfo{author}{\bibfnamefont{M.}~\bibnamefont{Palassini}} \bibnamefont{and}
  \bibinfo{author}{\bibfnamefont{S.}~\bibnamefont{Caracciolo}},
  \bibinfo{journal}{Phys. Rev. Lett.} \textbf{\bibinfo{volume}{82}},
  \bibinfo{pages}{5128} (\bibinfo{year}{1999}),
  \bibinfo{note}{cond-mat/9904246}.

\bibitem[{\citenamefont{Houdayer and
  Martin}(1999{\natexlab{b}})}]{HoudayerMartin99b}
\bibinfo{author}{\bibfnamefont{J.}~\bibnamefont{Houdayer}} \bibnamefont{and}
  \bibinfo{author}{\bibfnamefont{O.~C.} \bibnamefont{Martin}},
  \bibinfo{journal}{Phys. Rev. Lett.} \textbf{\bibinfo{volume}{83}},
  \bibinfo{pages}{1030} (\bibinfo{year}{1999}{\natexlab{b}}),
  \bibinfo{note}{cond-mat/9901276}.

\bibitem[{\citenamefont{Hukushima and Nemoto}(1996)}]{HukushimaNemoto96}
\bibinfo{author}{\bibfnamefont{K.}~\bibnamefont{Hukushima}} \bibnamefont{and}
  \bibinfo{author}{\bibfnamefont{K.}~\bibnamefont{Nemoto}},
  \bibinfo{journal}{J. Phys. Soc. Jpn.} \textbf{\bibinfo{volume}{65}},
  \bibinfo{pages}{1604} (\bibinfo{year}{1996}),
  \bibinfo{note}{cond-mat/9512035}.

\bibitem[{\citenamefont{Fisher and Huse}(1987)}]{FisherHuse87}
\bibinfo{author}{\bibfnamefont{D.~S.} \bibnamefont{Fisher}} \bibnamefont{and}
  \bibinfo{author}{\bibfnamefont{D.~A.} \bibnamefont{Huse}},
  \bibinfo{journal}{J. Phys. A Lett.} \textbf{\bibinfo{volume}{20}},
  \bibinfo{pages}{L1005} (\bibinfo{year}{1987}).

\bibitem[{\citenamefont{Krzakala and Parisi}(2003)}]{KrzakalaParisi03}
\bibinfo{author}{\bibfnamefont{F.}~\bibnamefont{Krzakala}} \bibnamefont{and}
  \bibinfo{author}{\bibfnamefont{G.}~\bibnamefont{Parisi}}
  (\bibinfo{year}{2003}), \bibinfo{note}{cond-mat/0304590}.

\end{thebibliography}

\end{document}